\documentclass[aps,prl,twocolumn,showpacs,amsmath,amssymb,superscriptaddress,10pt]{revtex4-1}

\usepackage[utf8]{inputenc}

\usepackage{graphicx}
\usepackage[colorlinks=true,citecolor=blue]{hyperref}
\usepackage{units}
\usepackage{color}

\usepackage{verbatim} 

\newcommand{\ket}[1]{|#1\rangle}
\renewcommand{\vec}[1]{\mathbf{#1}}

\DeclareMathOperator{\T}{T}

\def\up{\uparrow}
\def\down{\downarrow}

\begin{document}

\title{Unconventional Superconductivity in Double Quantum Dots}

\author{Bj\"orn Sothmann}
\affiliation{Département de Physique Théorique, Université de Genève, CH-1211 Genève 4, Switzerland}
\author{Stephan Weiss}
\affiliation{Theoretische Physik, Universit\"at Duisburg-Essen and CENIDE, 47048 Duisburg, Germany}
\author{Michele Governale}
\affiliation{School of Physical and Chemical Sciences and MacDiarmid Institute for Advanced Materials and Nanotechnology, Victoria University of Wellington, PO Box 600, Wellington 6140, New Zealand}
\author{J\"urgen K\"onig}
\affiliation{Theoretische Physik, Universit\"at Duisburg-Essen and CENIDE, 47048 Duisburg, Germany}

\date{\today}

\begin{abstract}
The formation of electron pairs is a prerequisite of superconductivity. 
The fermionic nature of electrons yields four classes of superconducting correlations with definite symmetry in spin, space and time.
Here, we suggest double quantum dots coupled to conventional $s$-wave superconductors in the presence of inhomogeneous magnetic fields as a model system exhibiting unconventional pairing.
Due to their small number of degrees of freedom, tunable by gate voltages, quantum-dot systems are ideal to gain fundamental insight in unconventional pairing.
We propose two detection schemes for unconventional superconductivity, based on either Josephson or Andreev spectroscopy. 
\end{abstract}


\maketitle

\paragraph{Introduction.--}
Conventional superconductivity is well understood in terms of Bardeen-Cooper-Schrieffer (BCS) theory by the formation of spin-singlet Cooper pairs of electrons~\cite{bardeen_theory_1957}.
In certain unconventional superconductors, spin-triplet pairs are involved~\cite{mackenzie_superconductivity_2003}.
A further generalization is the notion of pairing of electrons at different times. 
There are four different classes of superconducting correlations with definite symmetries in spin, space (momentum), and time (frequency) under exchange of two electrons forming a Cooper pair.
Even-frequency spin-singlet (odd in spin) Cooper pairs appear in conventional $s$-wave but also in high-$T_c$ $d-$wave~\cite{tsuei_pairing_2000} superconductors (even in space) while $p$- and $f$-wave pairing (odd in space) supports even-frequency spin triplets (even in spin).
There is experimental evidence for triplet $p$-wave pairing (similar to superfluid $^3$He~\cite{leggett_theoretical_1975}) in Sr$_2$RuO$_4$~\cite{mackenzie_superconductivity_2003}.
Recent proposals to induce triplet correlations in nanowires with strong spin-orbit interaction in proximity to $s$-wave superconductors were motivated by the prospect of creating Majorana fermions at the ends of the wire~\cite{lutchyn_majorana_2010,oreg_helical_2010}. 
By a similar mechanism, Majorana fermions can also be generated in double quantum dots~\cite{leijnse_parity_2012,sothmann_2013}.
The idea of odd-frequency pairing has first been brought up by Berezinskii~\cite{berezinskii_new_1974} as a possible explanation of superfluid $^3$He but has experienced a revival in the context of superconductor-ferromagnet heterostructures~\cite{bergeret_odd_2005}.
For noncollinear magnetizations, e.g., due to domain walls~\cite{kadigrobov_quantum_2001,bergeret_long-range_2001,fominov_josephson_2007,braude_fully_2007,volkov_odd_2008,cottet_inducing_2011,belzig_2011}, spin-active interfaces~\cite{eschrig_triplet_2008,linder_pairing_2009}, multiple noncollinear magnetized 
ferromagnetic layers~\cite{volkov_odd_2003,houzet_long_2007,belzig_2011}, 
helical magnets~\cite{volkov_helical_2006,champel_0-_2008,halasz_critical_2009}, or spin-orbit coupling \cite{bergeret_so},
odd-frequency correlations with finite spin polarization can penetrate deeply into ferromagnets, as confirmed in several experiments~\cite{sosnin_superconducting_2006,keizer_spin_2006,khaire_observation_2010,robinson_controlled_2010,sprungmann_evidence_2010,anwar_long-range_2010,klose_optimization_2012}.
Odd-frequency triplet pairing also appears in diffusive normal metals contacted by an even-frequency triplet superconductor \cite{tanaka_2007}.
Finally, odd-frequency singlet superconductivity has only been theoretically predicted~\cite{balatsky_new_1992} without experimental confirmation so far.

Quantum dots coupled to conventional superconductors show an interesting interplay of proximity effect and Coulomb interaction~\cite{de_franceschi_hybrid_2010,martin-rodero_josephson_2011}.
They have also been suggested as a tool to {\it detect} unconventional pairing \cite{tiwari_2013}.
In this Letter, we propose double quantum dots as an ideal system to {\it generate all four types} of superconducting correlations {\it in a single device} and to {\it control} them via gate and bias voltages and inhomogeneous magnetic fields.
This is a substantial advance over proposals which are specific to one type of correlations only, in which tuning is limited, or which describe equilibrium scenarios only.

\paragraph{Model.--}
\begin{figure}
	\includegraphics[width=\columnwidth]{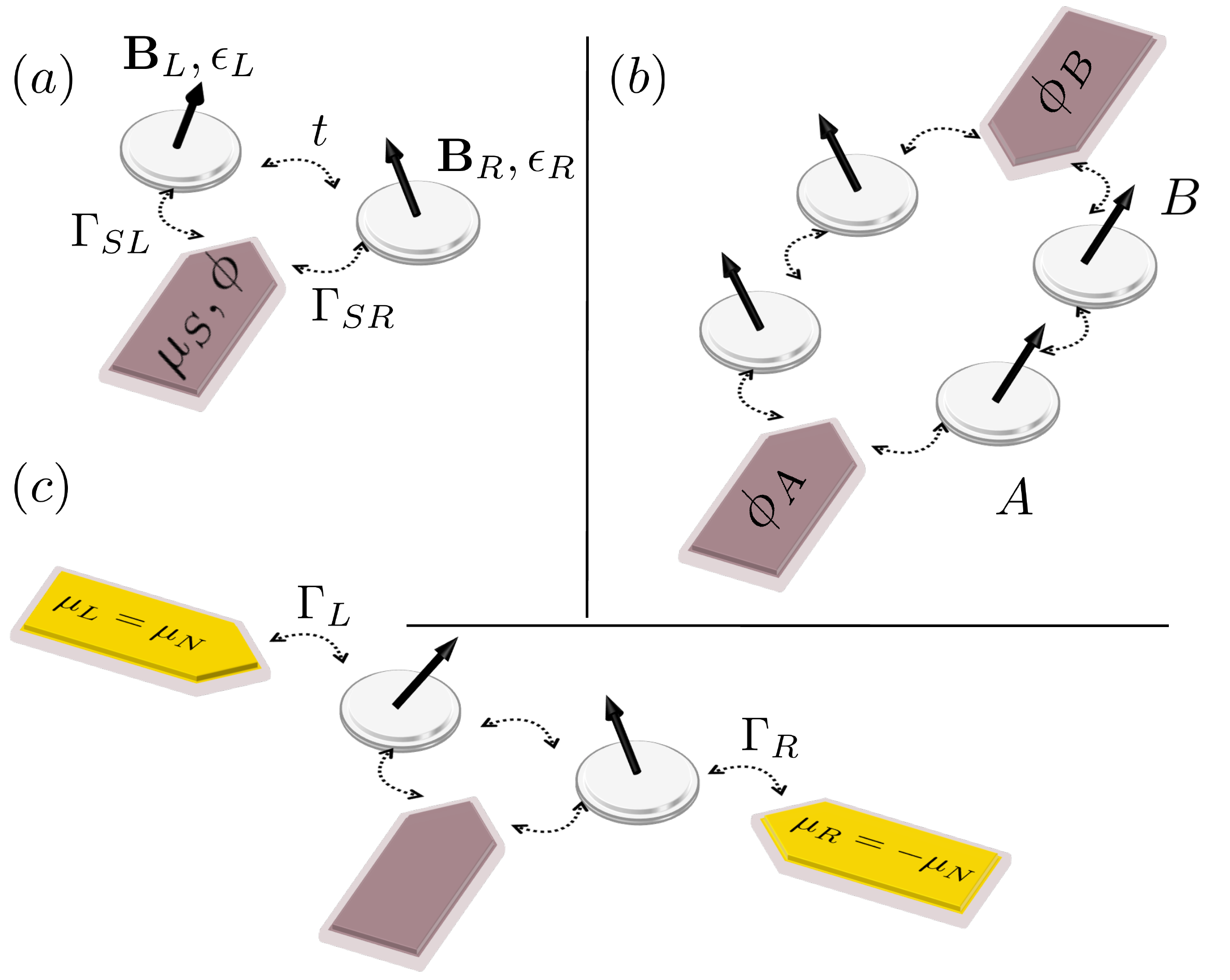}
	\caption{\label{fig:model}(Color online) (a) A double quantum dot in an inhomogeneous magnetic field and tunnel coupled to a superconducting lead yields unconventional pair amplitudes. 
	They can be probed via (b) the DC Josephson current through two coupled double quantum dots and (c) the Andreev current from the superconductor into two normal leads. 
	Dashed arrows denote tunnel couplings.
}
\end{figure}

We consider a DQD tunnel-coupled to a grounded BCS superconductor as indicated in Fig.~\ref{fig:model}(a). 
The effective double-dot Hamiltonian is $H_\text{DQD}=\sum_i H_i+H_\text{inter}+H_\text{prox}+H_\text{tun}$. 
Here, $H_i=\varepsilon_i\sum_\sigma n_{i\sigma}+\vec B_i\cdot\hat{\vec S}_i/\hbar+U_i n_{i\up}n_{i\down}$, where $n_{i\sigma}=c_{i\sigma}^\dagger c_{i\sigma}$, describes quantum dot $i=\text{L},\text{R}$ with a single level of energy $\varepsilon_i$, local magnetic field $\vec B_i$ (in units of $g\mu_\text{B}$) acting on dot spin $\hat{\vec S}_i=\hbar\sum_{\sigma\sigma'}c_{i\sigma}^\dagger\boldsymbol\sigma_{\sigma\sigma'}c_{i\sigma'}/2$, and intradot charging energy $U_i$. 
The interdot Coulomb interaction is given by $H_\text{inter}=U\sum_{\sigma\sigma'}n_{\text{L}\sigma}n_{\text{R}\sigma'}$. 
Superconducting correlations induced in the DQD via tunnel coupling to a superconducting lead are taken into account by the effective proximity Hamiltonian $H_\text{prox}=-\sum_i(\Gamma_{\text{S}i}/2)(c_{i\up}^\dagger c_{i\down}^\dagger+\text{H.c.})-(\Gamma_\text{S}/2)(c_{\text{R}\up}^\dagger c_{\text{L}\down}^\dagger-c_{\text{R}\down}^\dagger c_{\text{L}\up}^\dagger+\text{H.c.})$, which becomes exact in the limit of an infinite superconducting 
gap \cite{rozhkov_interacting-impurity_2000,eff-ham.tanaka,eff-ham.karrasch,eff-ham.meng,eldridge_superconducting_2010,futterer_2013}.
The strength of the local proximity effect is governed by the tunnel-coupling strength $\Gamma_{\text{S}i}=2\pi|t_i|^2\rho$ where $t_i$ and $\rho$ are the spin-independent tunnel amplitudes and density of states at the Fermi energy.
The strength of the nonlocal proximity effect $\Gamma_\text{S}$ may reach up to $\sqrt{\Gamma_{\text{SL}}\Gamma_{\text{SR}}}$ if both quantum dots are tunnel coupled to the same states in the superconducting leads, but is reduced otherwise.
A tunnel coupling between the dots is modeled by $H_\text{tun}=t\sum_{\sigma} (c_{\text{L}\sigma}^\dagger c_{\text{R}\sigma}+\text{H.c.})$. 
To address the situation $U_i \gg U$, it is sufficient to account for double occupancy of an individual dot only in virtual states.

\paragraph{Pair amplitude and order parameter.--}
To quantify  the strength of both conventional and unconventional superconducting correlations in the DQD with definite symmetry in spin, space and time, respectively, we form linear combinations of the pair amplitudes
\begin{equation}
	F_{i\sigma i'\sigma'}(t)=\langle\T c_{i\sigma}(t)c_{i'\sigma'}(0)\rangle \, ,
\end{equation}
where $\text{T}$ is the time-ordering operator. 
Non-local singlet correlations are characterized by $F_\text{e/o}^S=(F_{\text{L}\down\text{R}\up}-F_{\text{L}\up\text{R}\down}\mp F_{\text{R}\up\text{L}\down} \pm F_{\text{R}\down\text{L}\up})/(2\sqrt{2})$, non-local triplet by $F_\text{e/o}^{T^+}=(F_{\text{L}\up\text{R}\up}\mp F_{\text{R}\up\text{L}\up})/2$, 
$F_\text{e/o}^{T^0}=(F_{\text{L}\down\text{R}\up}+F_{\text{L}\up\text{R}\down}\mp F_{\text{R}\up\text{L}\down} \mp F_{\text{R}\down\text{L}\up})/(2\sqrt{2})$, and $F_\text{e/o}^{T^-}=(F_{\text{L}\down\text{R}\down}\mp F_{\text{R}\down\text{L}\down})/2$. 
The upper (lower) sign corresponds to even-(odd-)frequency pairing.
The prefactors are consistent with the definitions $\sqrt{2}\ket{S}=(c_{\text{R}\up}^\dagger c_{\text{L}\down}^\dagger-c_{\text{R}\down}^\dagger c_{\text{L}\up}^\dagger)\ket{0}$, $\ket{T^+}=c_{\text{R}\uparrow}^\dagger c_{\text{L}\uparrow}^\dagger\ket{0}$, $\sqrt{2}\ket{T^0}=(c_{\text{R}\up}^\dagger c_{\text{L}\down}^\dagger+c_{\text{R}\down}^\dagger c_{\text{L}\up}^\dagger)\ket{0}$, and $\ket{T^-}=c_{\text{R}\downarrow}^\dagger c_{\text{L}\downarrow}^\dagger\ket{0}$ for the singlet and triplet states, where $|0\rangle$ denotes an empty DQD.
The latter can be transformed into a cartesian vector via $|T_x\rangle=((|T^-\rangle-|T^+\rangle)/\sqrt{2}$, $|T_y\rangle=i(|T^-\rangle+|T^+\rangle)/\sqrt{2}$, and $|T_z\rangle=|T^0\rangle$. 
To define order parameters, we characterize the pair-amplitude functions by single numbers.
We use $\Delta_\text{e}=F_\text{e}(0)$ for even-frequency and $\Delta_\text{o}= \hbar (d F_\text{o}(t)/dt)|_{t=0}$ for odd-frequency pairing ~\cite{balatsky_even-_1993}.  
They form complex-valued scalars $\Delta_\text{e/o}^S$ and cartesian vectors $\boldsymbol \Delta_\text{e/o}^T$.  
In the absence of magnetic fields, $\boldsymbol \Delta_\text{e/o}^T = 0$ due to spin-rotation symmetry.
Similarly, for a collinear magnetic configuration, ${\bf B}_\text{L} \parallel {\bf B}_\text{R}$, only the triplet component along the symmetry axis is nonzero, $\boldsymbol \Delta_\text{e/o}^T \parallel {\bf B}_\text{L}$, i.e., only one of the triplet states can participate in superconducting correlations.  
The most interesting effects, however, appear for noncollinear magnetism, similar to SFS heterostructures~\cite{bergeret_odd_2005}.

\paragraph{Even-frequency pairing.--}
In the following, we restrict ourselves to the case $|{\bf B}_\text{L}|=|{\bf B}_\text{R}|$, for which the average magnetic field ${\bf B}=({\bf B}_\text{L}+{\bf B}_\text{R})/2$ and the difference $\Delta {\bf B}={\bf B}_\text{L}-{\bf B}_\text{R}$ are perpendicular to each other, i.e. the unit vectors along ${\bf B}$, $\Delta {\bf B}$, and ${\bf B} \times \Delta {\bf B}$ form an orthogonal coordinate system.
The presence of a superconducting lead couples $|0\rangle$ to $|S\rangle$, i.e., $\langle S|H_\text{DQD}|0\rangle =-\Gamma_\text{S}/\sqrt{2}$ generates a finite even-singlet order parameter $\Delta_\text{e}^S$.
In combination with $\langle T_{\Delta B}|H_\text{DQD}|S\rangle =-\Delta B/2$, this leads to a finite even-triplet order parameter $\Delta_\text{e}^{T_{\Delta B}}$ along the $\Delta {\bf B}$-direction.
Furthermore, $\langle T_{B\times\Delta B}|H_\text{DQD}|T_{\Delta B}\rangle =-iB$ yields a finite $\Delta_\text{e}^{T_{B\times \Delta B}}$, while $\Delta_\text{e}^{T_B}=0$ (for ${\boldsymbol \Delta}{\bf  B}\perp {\bf B}$).
A qualitative understanding of how the even-frequency order parameters depend on gate voltage (via the detuning $\delta=\epsilon_\text{L}+\epsilon_\text{R}+U$ between empty and doubly occupied DQD) and magnetic field (in particular the angle $\alpha = 2 \arctan (\Delta B/2B)$ enclosed by ${\bf B}_\text{L}$ and ${\bf B}_\text{R}$) can be obtained by ignoring all states with single occupation and mapping the resulting system onto an effective two-state model.
For this, we distinguish various limits.

(i) For $\Delta B\gg \Gamma_\text{S}$, we first diagonalize the Hamiltonian in the singlet-triplet subspace, and couple, then, the state with the lowest energy $\delta - \epsilon_\text{B}$ where $\epsilon_\text{B}=\sqrt{B^2+\Delta B^2/4}$ to the empty state.
The resulting ground state yields a resonant behavior of the order parameters
\begin{subequations}
\label{i}
\begin{align}
	\left|\Delta_\text{e}^S\right| 
	&=
	\frac{\Gamma_\text{S} \sin^2 (\alpha/2) }{2\sqrt{2}\sqrt{\Gamma_\text{S}^2\sin^2 (\alpha/2) +(\delta -\epsilon_\text{B})^2}}
\\
	\left|\Delta_\text{e}^{T_{\Delta B}}\right|
	&= \left|\Delta_\text{e}^S\right| / \sin (\alpha/2) 
\\
	\left|\Delta_\text{e}^{T_{B\times\Delta B}}\right|
	&= \left|\Delta_\text{e}^S\right| / \tan (\alpha/2) \, ,
\label{delBggGamS}
\end{align}
\end{subequations} 
as function of detuning $\delta$ with resonance position $\epsilon_\text{B}$ and width $\Gamma_\text{S}\sin(\alpha/2)$.

(ii) In the limit $\Gamma_\text{S}, B\gg \Delta B$, we take the lowest-energy eigenstates of the empty-singlet subspace and the triplet subspace, respectively.
The inhomogeneity $\Delta B$ then couples these states, which yields 
\begin{subequations}
\label{ii}
\begin{align}
	\left|\Delta_\text{e}^S\right|	
	&=
	\frac{\Gamma_\text{S}}{\sqrt{2}\sqrt{2\Gamma_\text{S}^2+\delta^2}}
	\label{DSii}
\\
	\left|\Delta_\text{e}^{T_{\Delta B}}\right| 
	&= 
	\frac{\left|\Delta_\text{e}^S\right| \Delta B /2}{\sqrt{(\delta+2\epsilon_\text{A}-2B)^2+(1-\frac{\delta}{2\epsilon_\text{A}})\Delta B^2}}
	\label{DTBDBii}
\\
	\left|\Delta_\text{e}^{T_{B\times\Delta B}}\right| 
	&= 
	\left|\Delta_\text{e}^{T_{\Delta B}}\right| \, ,
\end{align}
\end{subequations}
where we defined $2\epsilon_\text{A}=\sqrt{2\Gamma_\text{S}^2+\delta^2}$.
The resonance for the singlet order parameter is at $\delta = 0$ with width $\sqrt{2} \Gamma_\text{S}$. 
The extra factor that the triplet order parameters acquire, see Eq.~(\ref{DTBDBii}), displays a resonance at $\delta = \frac{B^2-\Gamma_\text{S}^2/2}{B} + {\cal O}(\Delta B^2)$ with a width that scales with $\Delta B$.

(iii) For $\Gamma_\text{S} \gg \Delta B \gtrsim B$ there is no effective two-state model since the triplet states are too close in energy to justify a truncation to the lowest-energy state only.
An exception is the special point $B=0$ (antiparallel magnetic fields), for which the triplet states perpendicular to $\Delta {\bf B}$ decouple from the rest of the Hamiltonian.
In that case, $|\Delta_\text{e}^S|$ is given by Eq.~(\ref{DSii}), and $\Delta_\text{e}^{T_B} = \Delta_\text{e}^{T_{B \times \Delta B}} = 0$,
\begin{equation}
\label{iii}
	\left|\Delta_\text{e}^{T_{\Delta B}}\right| 
	= 
	\frac{\left|\Delta_\text{e}^S\right| \Delta B }{\sqrt{(\delta+2\epsilon_\text{A})^2+2(1-\frac{\delta}{2\epsilon_\text{A}})\Delta B^2}} \, .
\end{equation}

\paragraph{Odd-frequency pairing.--}
The odd-frequency order parameters can be expressed in terms of even-frequency counterparts accompanied by local and non-local expectation values of charge and spin on the DQD.
We define $N^i_j=\sum_{\sigma}\langle c_{i\sigma}^\dagger c_{j\sigma}\rangle$ and $\vec S^i_j=\hbar\sum_{\sigma\sigma'}\langle c_{i\sigma}^\dagger\boldsymbol\sigma_{\sigma\sigma'} c_{j\sigma'}\rangle/2$ as well as $N_i=N^i_i$, $\Delta N = N_\text{L}-N_\text{R}$, ${\vec S}_i= {\vec S}_i^i$, and $\vec S=\vec S_\text{L}+\vec S_\text{R}$. The detuning of the two dot levels is $\Delta\varepsilon=\varepsilon_\text{L}-\varepsilon_\text{R}$.  
We obtain the general relations
\begin{subequations}
\begin{align}
	\label{eq:OPOT}\boldsymbol{\Delta}_\text{o}^T&=-\frac{i}{2}\Delta\varepsilon\boldsymbol{\Delta}_\text{e}^T+\frac{i}{2}\vec B\Delta_\text{e}^S+\frac{1}{4}\Delta\vec B\times\boldsymbol{\Delta}_\text{e}^T\nonumber\\&\phantom{=} 	+\frac{i}{2\sqrt{2}\hbar} \left( \Gamma_\text{S} \vec S - \Gamma_\text{SL}\vec S^\text{L}_\text{R}-\Gamma_\text{SR}\vec S^\text{R}_\text{L}\right),
	\\
	\label{eq:OPOS}\Delta_\text{o}^S&=-\frac{i}{2}\Delta\varepsilon \Delta_\text{e}^S+\frac{i}{2}\vec B\cdot\boldsymbol{\Delta}_\text{e}^T
	\nonumber\\
	&\phantom{=} -\frac{i}{4\sqrt{2}} \left( \Gamma_\text{S} \Delta N +\Gamma_\text{SL} N^\text{L}_\text{R}-\Gamma_\text{SR}N^\text{R}_\text{L}\right)
\, .
\end{align}
\end{subequations}
They demonstrate the rich diversity of possible routes towards odd-frequency correlations in quantum-dot systems that may be absent in other systems.
Odd-triplet correlations coexist with even-singlet correlations in the presence of a magnetic field ${\bf B}$ and with even-triplet correlations for either finite level detuning $\Delta\varepsilon$ or finite inhomogeneity $\Delta {\bf B}$.
Furthermore, odd-triplet correlations are induced by a finite spin polarization on the DQD due to Zeeman splitting, FM contacts or spin orbit coupling. 
The odd-triplet state is unitary \cite{sigrist_phenomenological_1991} when induced by finite spin polarization, but is, in general, non-unitary in the other cases. 
Odd-singlet correlations coexist with even-singlet correlations for finite $\Delta\varepsilon$ and with even-triplet correlations for finite ${\bf B}$.
In addition, odd-singlet correlations may exist as a consequence of a finite local or non-local charge asymmetry.

\paragraph{Transport signatures.--}
In the following we propose two schemes to identify superconducting triplet correlations in a transport measurement \cite{footnote}.
The first setup consists of two DQDs $a=A,B$, see Fig.~\ref{fig:model}(b), each coupled to a superconductor with phase $\phi_a$.
To define the simplest possible model, we assume identical copies of $H_{{\rm DQD},a}$ except for different phases $\phi_a$ that need to be included in the tunnel coupling to the superconductors \cite{governale_real-time_2008}.
Furthermore, we set $H_\text{tun}=0$ within each DQD, and assume identical interdot Coulomb-interaction strengths $U$ between any pair of quantum dots. 
The two DQDs are connected to each other by tunnel couplings $H_\text{coupl}=\sum_{i\sigma} t_\sigma c_{i\sigma,A}^\dagger c_{i\sigma,B}+\text{H.c.}$
We consider the regime $\Gamma_{\text{S}i,a}\gg t_\sigma$, in which intra-DQD correlations dominate over inter-DQD correlations. 
However, we emphasize that all our conclusions stay valid for arbitrary values of inter- and intra-DQD couplings.

\begin{figure}
	\includegraphics[width=\columnwidth]{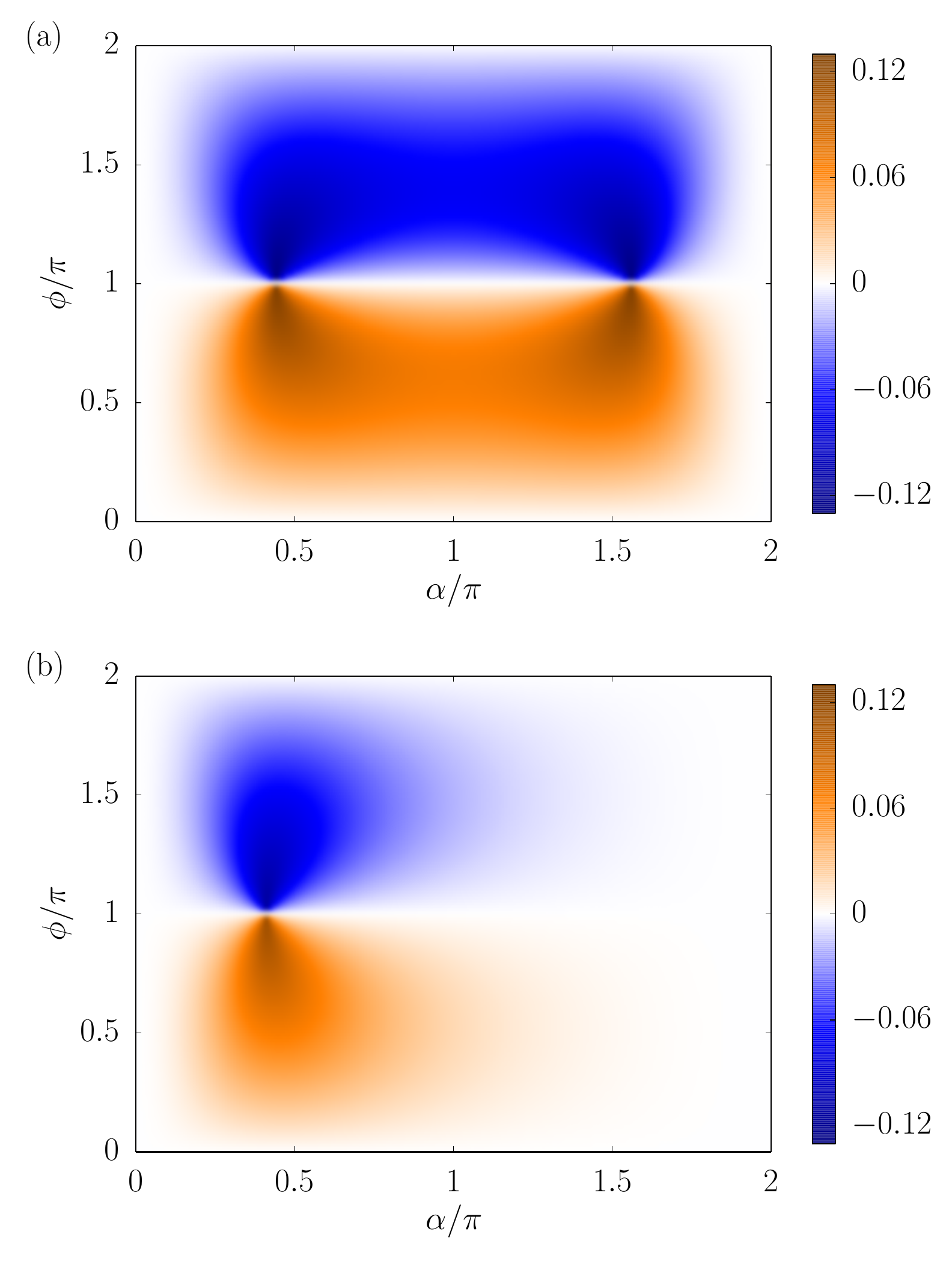}
	\caption{\label{fig:Josephson2} (Color online) Josephson current in units of $e\Gamma_{\text{S}i,a}/\hbar$ for $\Gamma_{\text{S}i,a}=U$, $\varepsilon_{i,a}= 3.5 U$, $|\vec B_{i,a}|=10U$, $\alpha_a=\alpha$, $k_\text{B} T=0$ for (a) spin-insensitive tunneling $t_\up=t_\down=0.1U$ and (b) fully spin-selective tunneling $t_{\up}=0.1U$, $t_\down=0$.
	The existence of a maximum of the Josephson current for an angle $\alpha_\text{max}$ with $0 < \alpha_\text{max} < \pi$ (that is for noncollinear magnetic fields) signals the presence of triplet correlations in the system.
}
\end{figure}

The zero-temperature Josephson current $J=(2e/\hbar)dE_0/d\phi$ between the two superconductors is given by the derivative of the ground-state energy $E_0$ with respect to the phase difference $\phi=\phi_A-\phi_B$.
We first consider the case of spin-insensitive tunneling, $t_\up=t_\down$.
As shown in Fig.~\ref{fig:Josephson2}(a), the Josephson current is suppressed for $\alpha=0$ (homogeneous magnetic field), increases with $\alpha$, reaches a maximum at some angle $0 < \alpha_\text{max} < \pi$, and then decreases up to the local minimum at $\alpha=\pi$ (antiparallel magnetic fields).
This behavior {\it cannot} be explained with superconducting singlet correlations alone.
Indeed, the notion that singlet pairs are energetically suppressed in a homogeneous magnetic field and that this suppression weakens with increasing $\alpha$ may be compatible with the {\it increase} of $J$ for $\alpha < \alpha_\text{max}$.
The very existence of a {\it maximum} at an angle $0 < \alpha_\text{max} < \pi$ describing a {\it noncollinear} magnetic field, however, is a clear signature of the presence of superconducting (even- and/or odd-) triplet correlations.
The existence of triplet correlations is even more evident for fully spin-selective tunnel barriers, see Fig.~\ref{fig:Josephson2}(b), where we chose $t_\down=0$ (for spin quantization axis along $\bf B$).
Since only spin-$\uparrow$ electrons can be transferred between the DQDs, any finite Josephson current {\it must} be due to transfer of {\it triplet} pairs.

The second proposal relies on measuring the Andreev current between normal and superconducting leads in a setup  shown in Fig.~\ref{fig:model}(c), similar to those experimentally realized in Refs.~\onlinecite{hofstetter_2009,hermann_2010}.
The additional, normal conducting leads, $H_\text{leads}=\sum_{k r \sigma}\epsilon_{k\sigma}a^\dag_{kr\sigma}a_{kr\sigma}$, are weakly coupled to the DQD by tunneling, $H_{T}=\sum_{kr\sigma} (t_r a_{kr\sigma}c^\dag_{r\sigma}+\text{H.c.})$ with $r=\text{L,R}$.
A bias voltage is antisymmetrically applied between normal and superconductors (electrochemical potentials $\mu_\text{L}=-\mu_\text{R}\equiv\mu_\text{N} = eV$ relative to the superconductor, $\mu_\text{S}=0$), and the (Andreev) current $I$ in the superconductor is measured.
We calculate the latter to first order in the tunnel-coupling strengths $\Gamma_\text{L/R}=2\pi\rho|t_\text{L/R}|^2$ (where $\rho$ is the density of states in the normal leads at the Fermi energy) in the wide-band limit by making use of a real-time diagrammatic approach for quantum-dot systems that involve superconducting leads~\cite{governale_real-time_2008,sothmann_probing_2010}. 
In the following, we consider the symmetric case, $\Gamma_\text{L}=\Gamma_\text{R}\equiv \Gamma_\text{N}$ and $\epsilon_\text{L}=\epsilon_\text{R}$. 

\begin{figure}
	\includegraphics[width=\columnwidth]{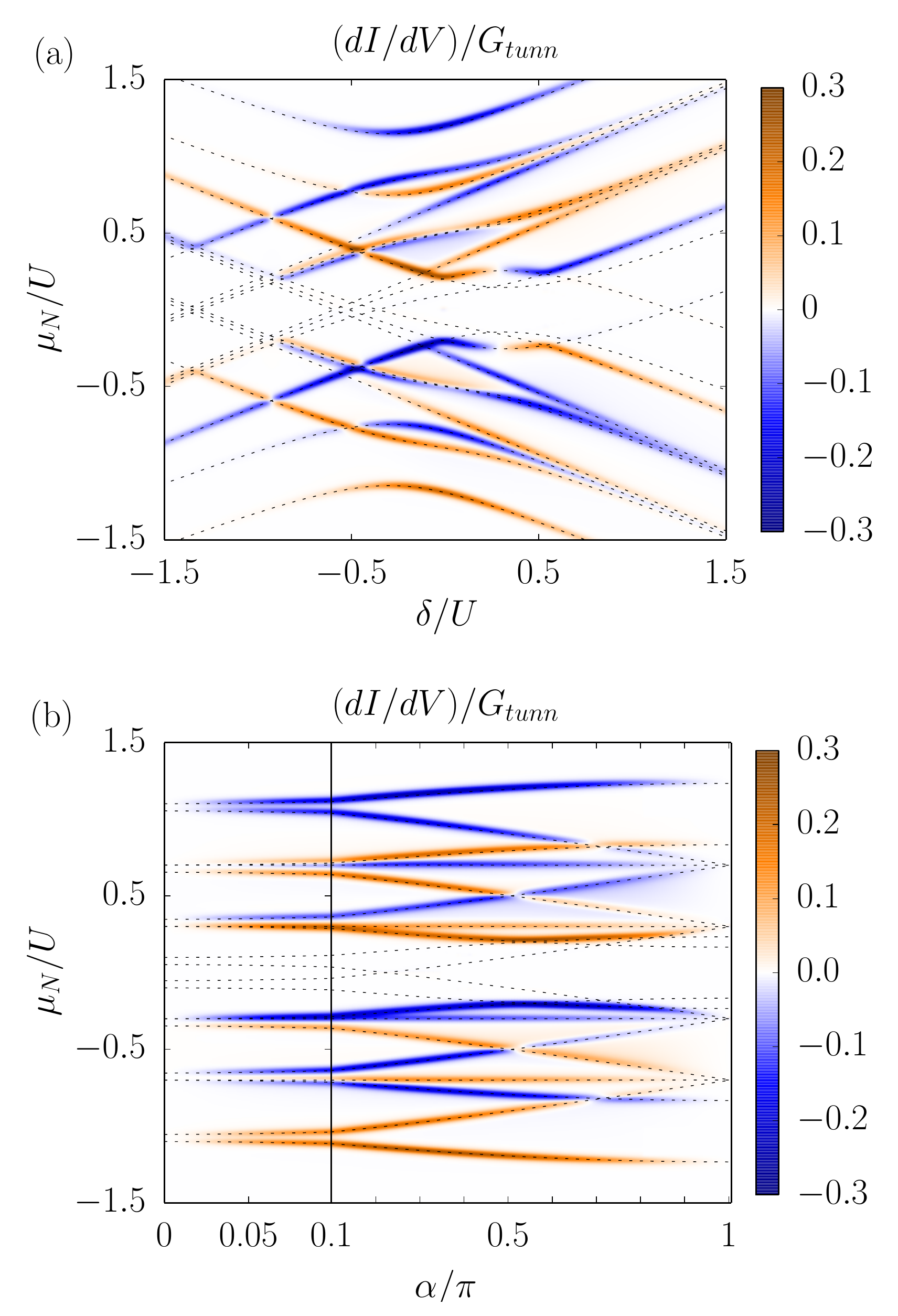}
	\caption{\label{fig:andreev} (Color online) Differential Andreev conductance in units of $G_\text{tunn}=(e^2/h)\Gamma_\text{N}/(2\pi k_\text{B}T)$ as a function of bias voltage $\mu_\text{N}$ and (a) detuning $\delta$ for $\alpha=\pi/2$ and (b) $\alpha$ for $\delta=0$.
	The chosen parameters are $\Gamma_\text{S}=U/2, |{\bf B}_\text{L/R}|=0.4U$, $t=0.001 U$, and $k_\text{B}T=0.01 U$.
	The nonmonotonic behaviour of the differential conductance on $\alpha$, resulting in a suppression around $\alpha=\pi$, is a signature of triplet correlations.}
\end{figure}

In Fig.~\ref{fig:andreev}(a), we show the differential conductance $dI/d V$ as a function of gate and bias voltage, $\delta$ and $\mu_\text{N}$, for $\alpha=\pi/2$. 
Resonances as a function of $\mu_\text{N}$ appear at the Andreev addition energies $\pm(E_{\text{even}}-E_{\text{odd}})$, shown as dashed lines, where $E_{\text{even/odd}}$ are eigenvalues of $H_\text{DQD}$ in the subspaces with an even/odd number of electrons in the DQD, respectively. 
The suppressed conductance for small $\mu_\text{N}$ is due to Coulomb blockade.
For large detunings $|\delta|$, the Andreev current decreases due to a suppression of the superconducting order parameters.
Similar as for the Josephson current discussed above, the presence of superconducting triplet correlations can be identified in the Andreev current by analyzing its dependence on the noncollinearity $\alpha$ of the magnetic field, see Fig.~\ref{fig:andreev}(b).
We find that transport is suppressed for $\alpha=0$, first increases but then {\it decreases} as a function of $\alpha$, and is suppressed again for $\alpha=\pi$.
The very fact that Andreev transport becomes maximal for {\it noncollinear} magnetic fields proves the presence of triplet correlations.
We have verified that this behavior survives for asymmetric tunnel couplings $\Gamma_\text{L} \neq \Gamma_\text{R}$ and other choices of parameters \cite{supp}.

\paragraph{Conclusions.--}
We demonstrate that unconventional superconducting correlations generically occur in DQDs coupled to a conventional superconductor and subject to an inhomogeneous magnetic field. 
To quantify them, we define order parameters for even-/odd-frequency singlet/triplet correlations and discuss under which conditions they occur.
Their dependence on gate and bias voltage as well as noncollinearity of the magnetic field can be probed both in Josephson and in Andreev transport spectroscopy.
We propose two measurement schemes to identify superconducting triplet correlations.
The distinctive feature of the proposed devices is that they simultaneously accommodate all possible types of pairing, which can be controlled via applied voltages and fields.

\paragraph{Acknowledgements.--}
We acknowledge financial support from the Swiss NSF and the NCCR QSIT and the DFG under project KO $1987/5$. Fruitful discussion with F.S. Bergeret and U. Z\"ulicke is acknowledged.

\pagebreak

\widetext

\renewcommand{\thefigure}{S\arabic{figure}}
\setcounter{figure}{0}

\begin{center}

{\large \bf Supplemental Material for: \smallskip \\ 
Unconventional Superconductivity in Double Quantum Dots}
\\

\bigskip 

Bj\"orn Sothmann, Stephan Weiss, Michele Governale, and J\"urgen K\"onig

\bigskip

\end{center}

The results presented in the main text and the conclusions drawn are not limited to specific choices of system parameters but rather generic.
To prove this, we show in the following results obtained for different parameters than those chosen in Figs.~2 and 3.
Furthermore, we plot the amplitudes of all superconducting order parameters for the Andreev-current setup.

\section{I. Josephson spectroscopy}

\begin{figure}[ht!]
\begin{minipage}[b]{0.49\textwidth}
	\includegraphics[width=\columnwidth]{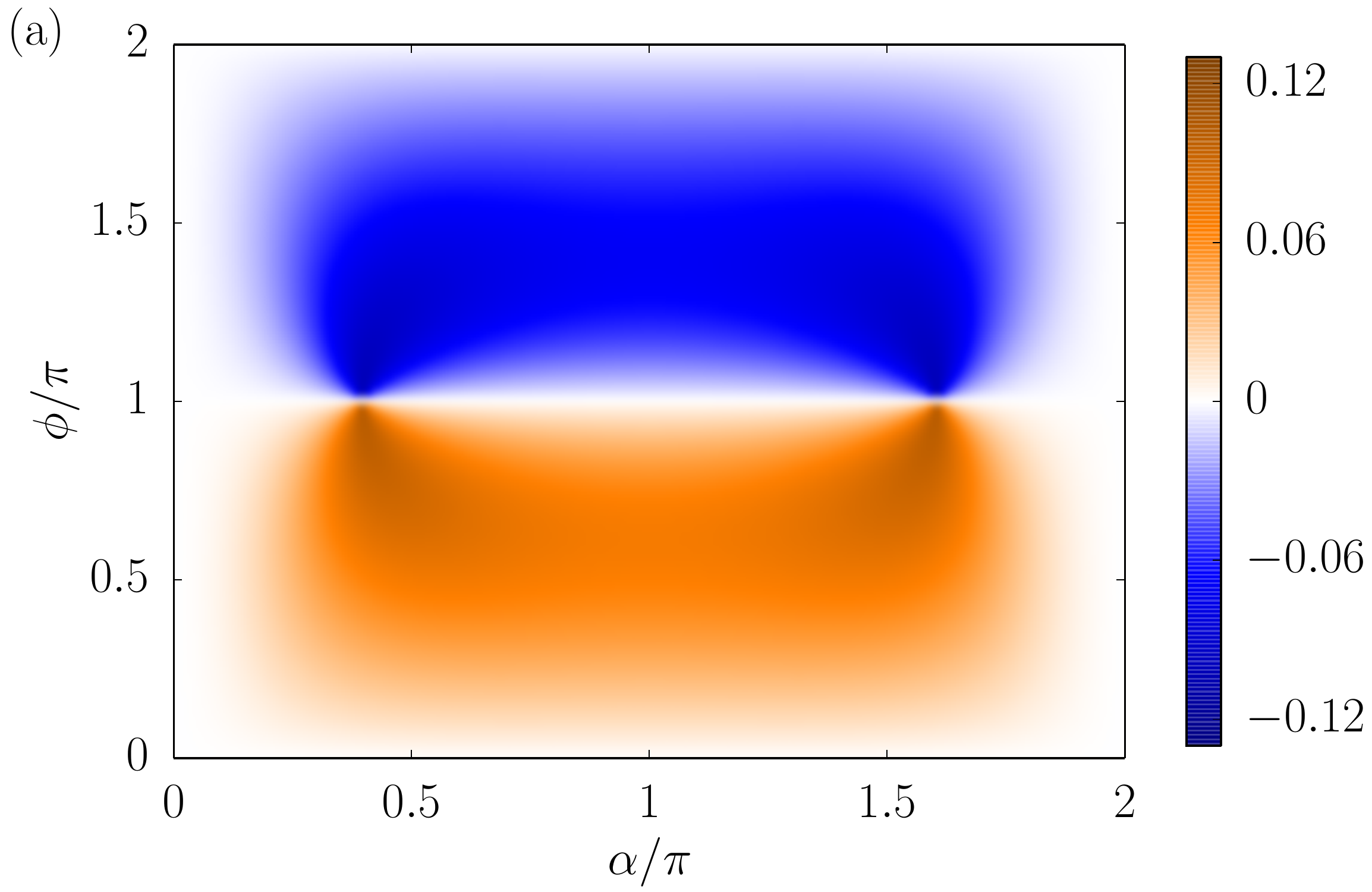}
	\end{minipage}
\begin{minipage}[b]{0.49\textwidth}
\includegraphics[width=\columnwidth]{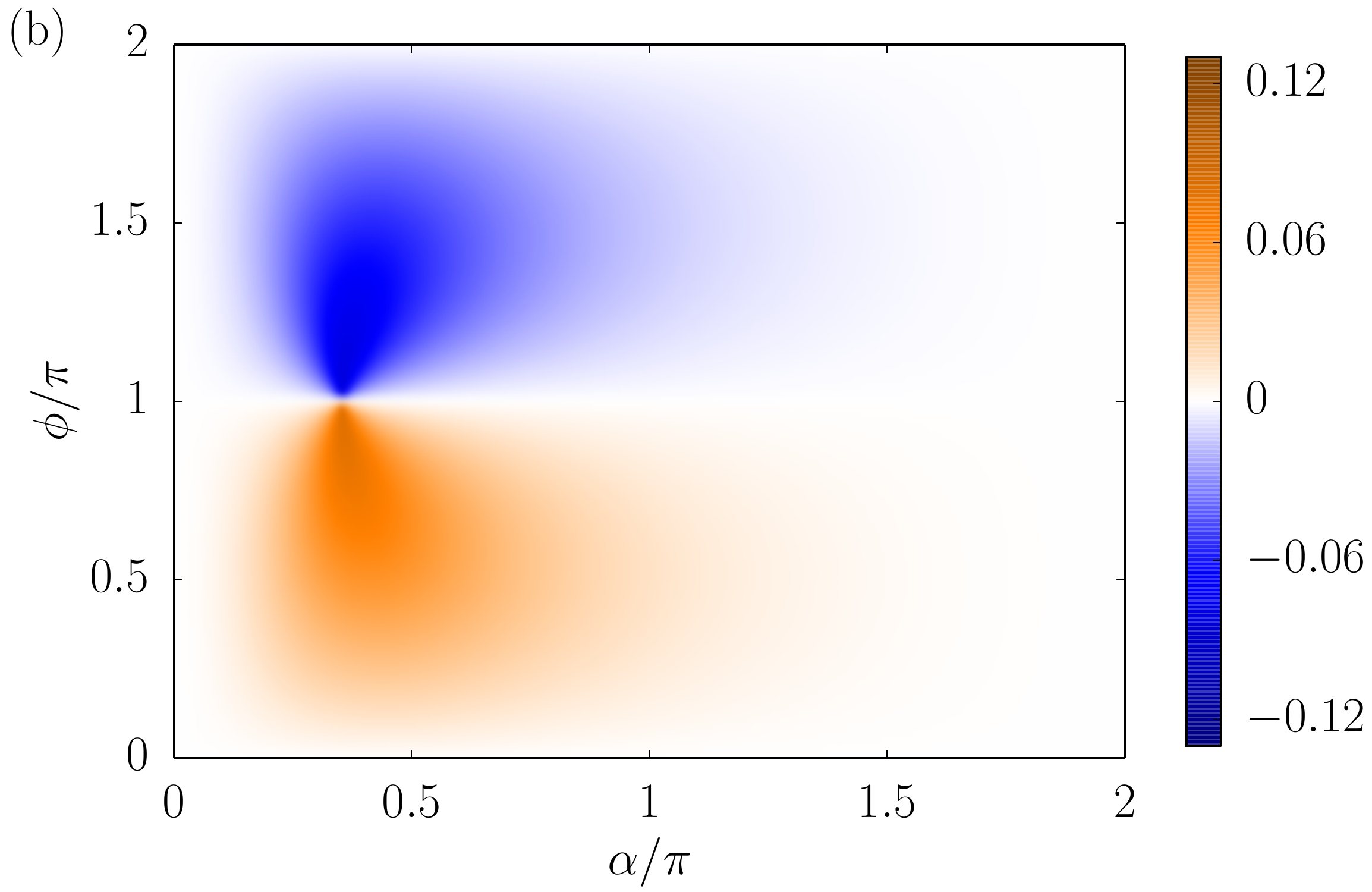}
	\end{minipage}
\caption{\label{fig:Josephson_cross}Josephson current for the same system as in Fig.~2 but now including tunneling between all dot-dot pairs with tunneling amplitude $t$ for dot-dot nearest neighbors and $t'=t/2$ for cross tunneling.}
\end{figure}

In the calculation for the Josephson current shown in Fig.~2, we assumed that only two out of the six possible dot-dot tunnel couplings are nonzero (shown as arrows in Fig.~1(b)).
Now, we compare this with the case that all six dot-dot couplings are present, see Fig.~\ref{fig:Josephson_cross}.
We take the same tunneling amplitude $t$ for all four dot-dot nearest neighbors, and a smaller amplitude $t'=t/2$ for the remaining two cross-tunneling terms.
The result is almost identical to Fig.~1. 
The only difference is a slight shift of the maximum position $\alpha_\text{max}$.
The presence of triplet correlations is still clearly visible.

\section{II. Andreev spectroscopy}

We now turn to Andreev spectroscopy for the device shown in Fig.~1(c).

\subsection{A. Weak tunnel coupling to the superconducting lead}

\begin{figure}[ht!]
\begin{minipage}[b]{0.47\textwidth}
	\includegraphics[width=\columnwidth]{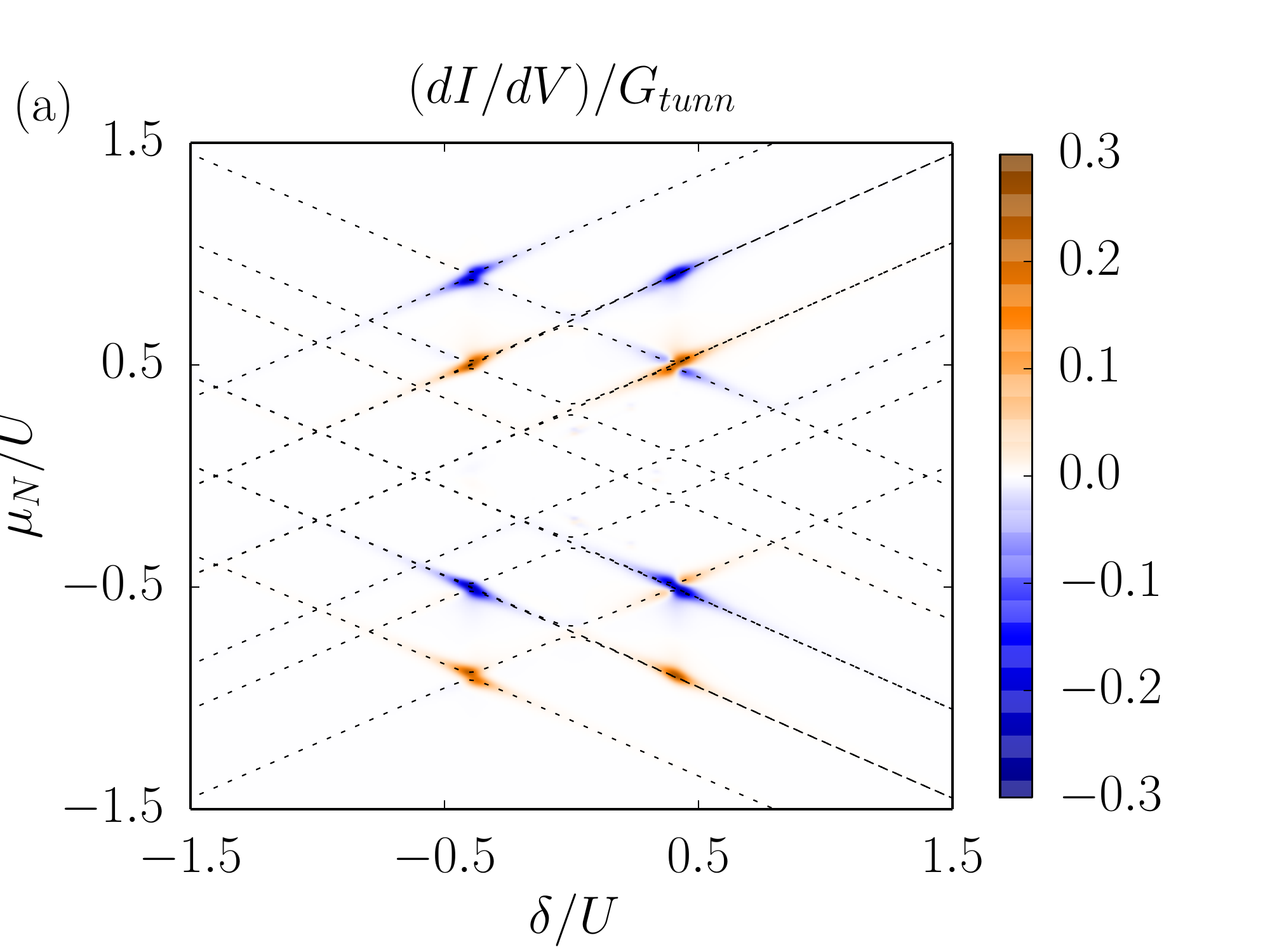}
\end{minipage}
\begin{minipage}[b]{0.47\textwidth}
	\includegraphics[width=\columnwidth]{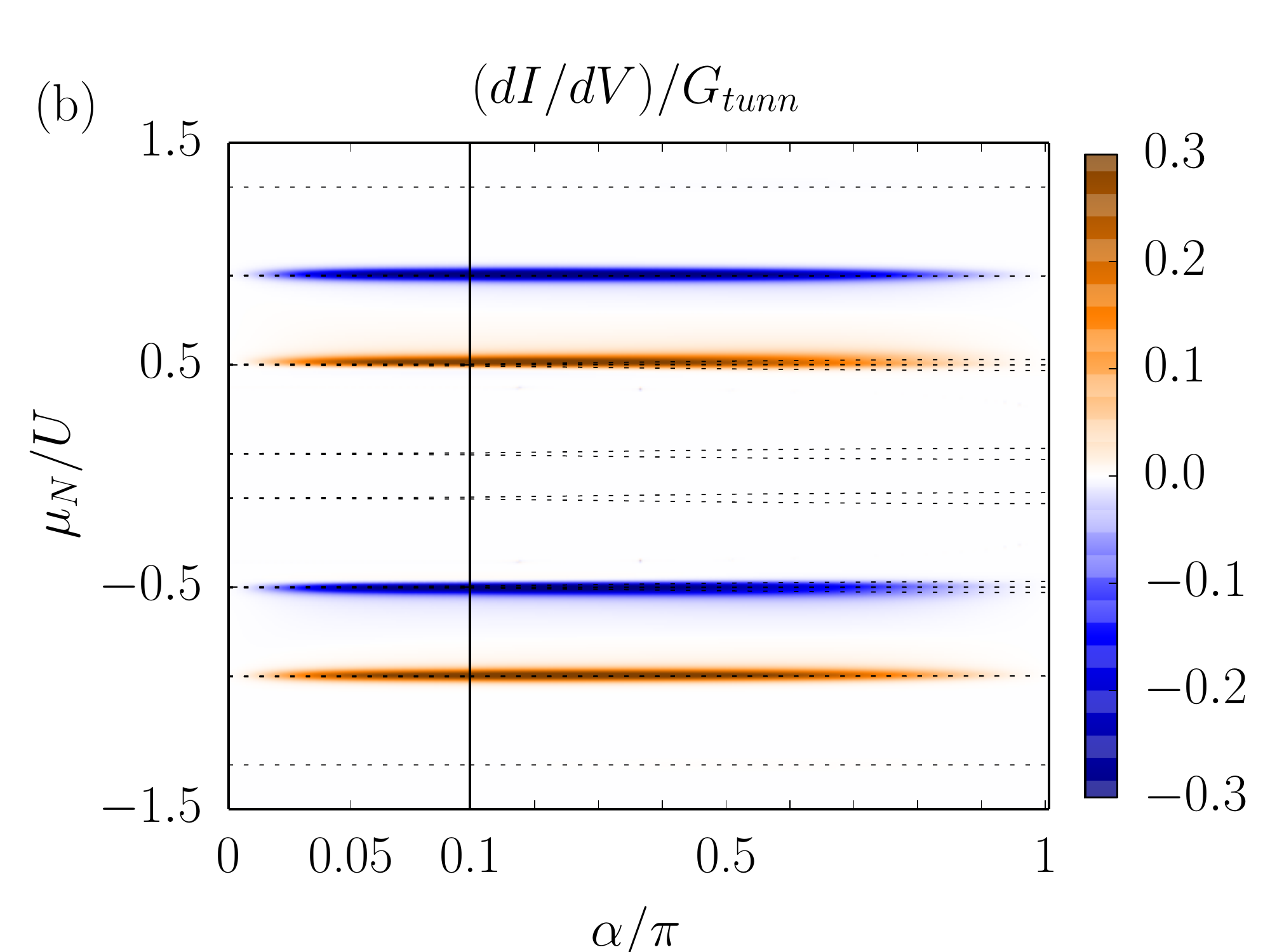}
	\end{minipage}
\caption{\label{fig:Andreev_small_gamS}Differential Andreev conductance for the same system as in Fig.~3 but for a weaker coupling to the superconductor, $\Gamma_\text{S}=0.05U$ and, in panel (b), a different choice of $\delta=0.4U$ in order to match the resonance condition. Again, the nonmonotonic dependence on $\alpha$ is a signature of superconducting triplet correlations.
}
\end{figure}

In the calculation for Fig.~3, we assumed a rather large coupling $\Gamma_\text{S}$ between double quantum dot and superconducting lead.
As a consequence, we got a strong proxmity effect with pronounced anticrossings in the Andreev addition energies.
Superconducting triplet correlations are, however, also visible in the limit of weak coupling, $\Gamma_\text{S}$, as shown in Fig.~\ref{fig:Andreev_small_gamS}, where we took the same parameters as in Fig.~3 but reduced $\Gamma_\text{S}$ by a factor of $10$.
In this regime, a perturbative treatment of $\Gamma_\text{S}$ to lowest order would be sufficient.
The Andreev addition spectrum is, then, independent of $\alpha$, Fig.~\ref{fig:Andreev_small_gamS}(b), and the anticrossings are no longer resolved, Fig.~\ref{fig:Andreev_small_gamS}(a).
Nevertheless, superconducting triplet correlations are clearly indicated by the nonmonontic behavior of the differential conductance as function of $\alpha$, taken at $\delta=0.4 U$ in Fig.~\ref{fig:Andreev_small_gamS}(b).
For this choice of $\delta$, we meet the resonance condition $\delta=\epsilon_\text{B}$ derived in Eqs.~(2a)-(2c) for the limit $\Delta B\gg \Gamma_\text{S}$, which is satisfied here.
Interestingly, we find in Fig.~\ref{fig:Andreev_small_gamS}(a) not only resonances at $\delta=\epsilon_\text{B}$ but also at $\delta=-\epsilon_\text{B}$.
The latter can be similarly derived as the former by diagonalizing the Hamiltonian in the singlet-triplet subspace but then selecting the state with the {\it highest} energy $\delta+\epsilon_\text{B}$, coupling it to the empty state, and, finally picking the lower-energy state of this effective two-state system.
This state is obviously not the ground state but can, nevertheless, be accessed at finite bias voltage.

\vspace*{-.5cm}

\subsection{B. Varying a global magnetic field in the presence of a fixed inhomogeneous one}

In the main text, we discussed the magnetic-field dependence by keeping the magnitudes of the local fields equal and fixed, $|{\bf B}_\text{L}|=|{\bf B}_\text{R}|$, and varying the angle $\alpha$ between their directions. 
Experimentally, it may be easier to generate non-collinear magnetic fields by varying a global (homogeneous) external magnetic field that is superimposed with a fixed inhomogeneous one.
To be specific, we choose ${\bf B}_\text{L}={\bf B}_\text{i}+{\bf B}_\text{g}$ and ${\bf B}_\text{R}=-{\bf B}_\text{i}+{\bf B}_\text{g}$ with an angle $\gamma$ 
 between ${\bf B}_\text{g}$ and ${\bf B}_\text{i}$.
The variation of $B_\text{g}$ will lead to a crossover from a nearly antiparallel (for $B_\text{g} \ll B_\text{i}$) to a noncollinear (for $B_\text{g} \sim B_\text{i}$) and, then, to a nearly parallel (for $B_\text{g} \gg B_\text{i}$) configuration.
In such a scenario, the local fields will, in general, differ in magnitude, $|{\bf B}_\text{L}|\neq |{\bf B}_\text{R}|$, such that now all three triplet states can couple to the empty state.
The only exception is the range of small angles $\gamma$ that satisfy $B_\text{g} \sin \gamma \ll t$.
In this case, the interdot tunnel coupling $t$ dominates over the left-right asymmetry introduced into the spectrum by ${\bf B}_\text{g}$, which yields $|{\bf B}_\text{L}| \approx |{\bf B}_\text{R}|$ (and one triplet state decouples from the other doubly-occupied states).

\begin{figure}[ht!]
\includegraphics[width=0.98\columnwidth]{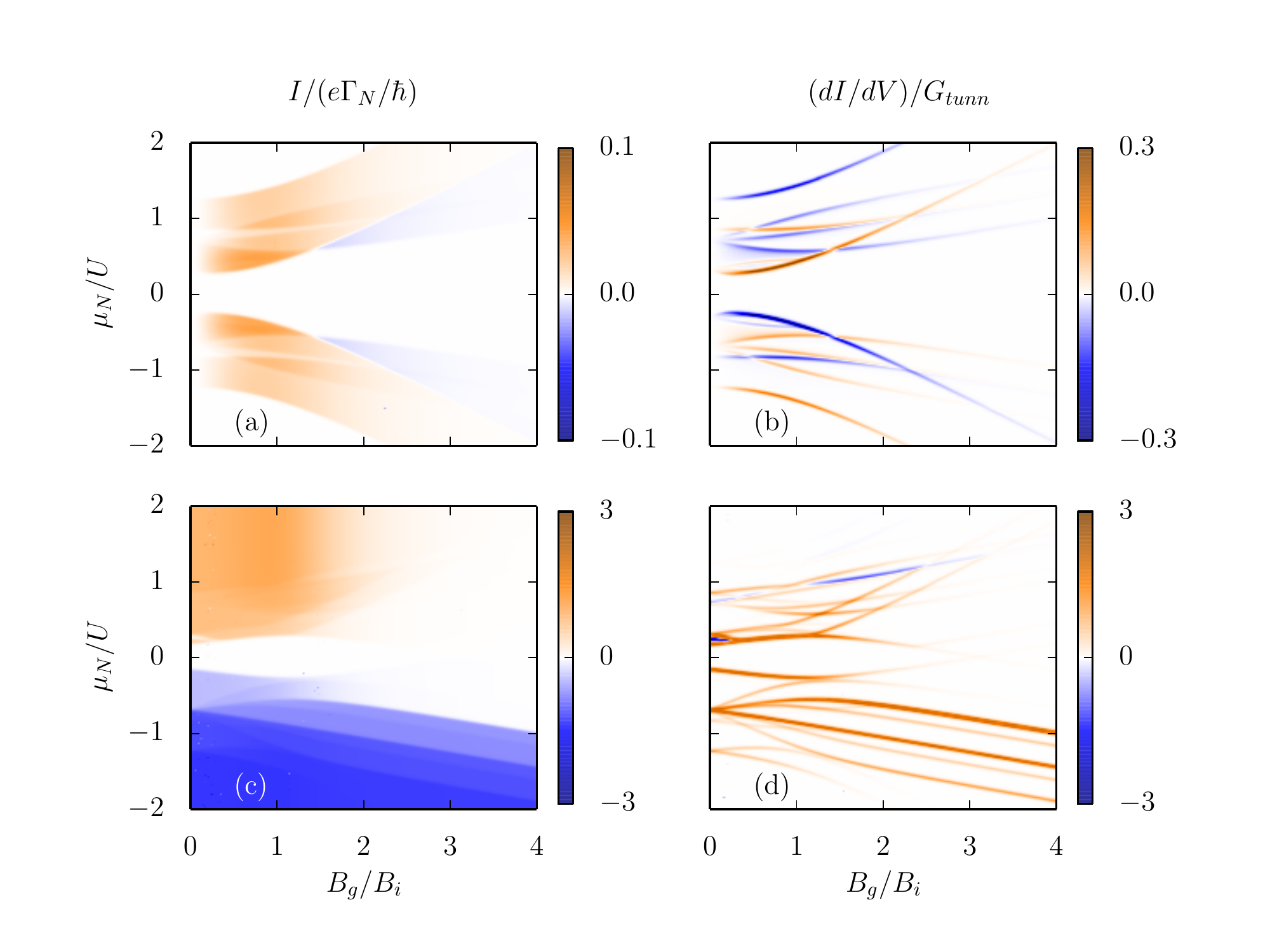}
\caption{\label{fig:Andreev_delB} (a) Andreev current and (b) differential conductance for antisymmetrically and (c) Andreev current and (d) differential conductance for symmetrically applied bias voltage.
The parameters are the same as in Fig.~3 with $B_\text{i}=0.4 U$ as well as $\gamma=0$ in (a) and (b) while $\gamma= \pi/4$ in (c) and (d). 
}
\end{figure}

In Fig.~\ref{fig:Andreev_delB} we show the dependence of (a) the current and (b) the conductance on $B_\text{g}$ for $\gamma=0$.
We find that in the regimes of nearly parallel ($B_\text{g} \gg B_\text{i}$) and nearly antiparallel ($B_\text{g} \ll B_\text{i}$) magnetic configuration, the current is suppressed as compared to the noncollinear case ($B_\text{g} \sim B_\text{i}$), which is a clear signature of triplet correlations, in agreement with what we find in Fig.~3.

\subsection{C. Symmetrically applied bias voltage}

The differential conductance shown in Fig.~3 were for the three-terminal device shown in Fig.~1(c) with {\it antisymmetrically} applied bias voltage, $\mu_\text{L} = - \mu_\text{R} = \mu_\text{N} = eV$ relative to the superconductor $\mu_\text{S}=0$. 
We now consider the case of a {\it symmetrically} applied bias voltage, $\mu_\text{L} = \mu_\text{R} = \mu_\text{N} = eV$, i.e., there is only a bias between normal leads and superconductor but not between the two normal leads which are short-cut, making the system effectively a two-terminal device.

The case of an antisymmetrically applied bias voltage has the advantage that in the absence of a noncollinear magnetic field transport is suppressed, which is interpreted as a signature of triplet correlations.
For a symmetrically applied bias voltage, the situation is different.
In this case, transport occurs already in the absence of any magnetic field [42].
There is, however, a special feature that still allows for the identification of triplet correlations.
As discussed in Ref.~[42], transport becomes blocked for bias voltages such that the electrons entering the double dot from the normal leads cannot get back but have to go into the superconductor.
Since two electrons entering the double dot from the normal leads are with finite probability in a triplet state, they cannot enter the superconductor (triplet blockade) unless superconducting triplet correlations are induced.
The triplet blockade leads to an absence of Andreev current for large and positive $\mu_\text{N}$ in the regimes of almost collinear magnetic fields, $B_\text{g} \gg B_\text{i}$ and $B_\text{g} \ll B_\text{i}$.
In the regime $B_\text{g} \sim B_\text{i}$, however, superconducting triplet correlations are induced and the triplet blockade is lifted.
The triplet blockade for $B_\text{g} \gg B_\text{i}$ is clearly visible in Fig.~\ref{fig:Andreev_delB}(c) and (d).
For $B_\text{g} \sim B_\text{i}$, the triplet blockade is lifted (we choose the angle $\gamma$ large enough such that $B_\text{g} \sin \gamma \gtrsim t$ guarantees that {\it all} triplet states can couple to the empty state in the regime $B_\text{g} \sim B_\text{i}$).
For the chosen parameters, a pronounced triplet blockade around $B_\text{g}=0$ occurs only for a small range that is not resolved in Fig.~\ref{fig:Andreev_delB}(c) and (d).
For negative $\mu_\text{N}$, the electrons are transfered from the superconductor to the normal leads, and triplet blockade does not appear for any value of $B_\text{g}$.

\section{III. Superconducting order parameters}
 
Andreev spectrocopy as shown in Fig.~3 gives an indirect access to the superconducting order parameters. 
From the $\alpha$-dependence of the conductance, we could deduce the presence of superconducting triplet correlations without having detailed information about the relative importance of the various superconducting order parameter as a function gate and bias voltage.
In the following, we provide this information by plotting the absolute values of the complex scalars $\Delta_\text{e/o}^S$ and vectors 
$\boldsymbol{\Delta}_\text{e/o}^T$.

\subsection{A. No detuning between dot levels}

\begin{figure}[ht!]
\includegraphics[width=0.98\columnwidth]{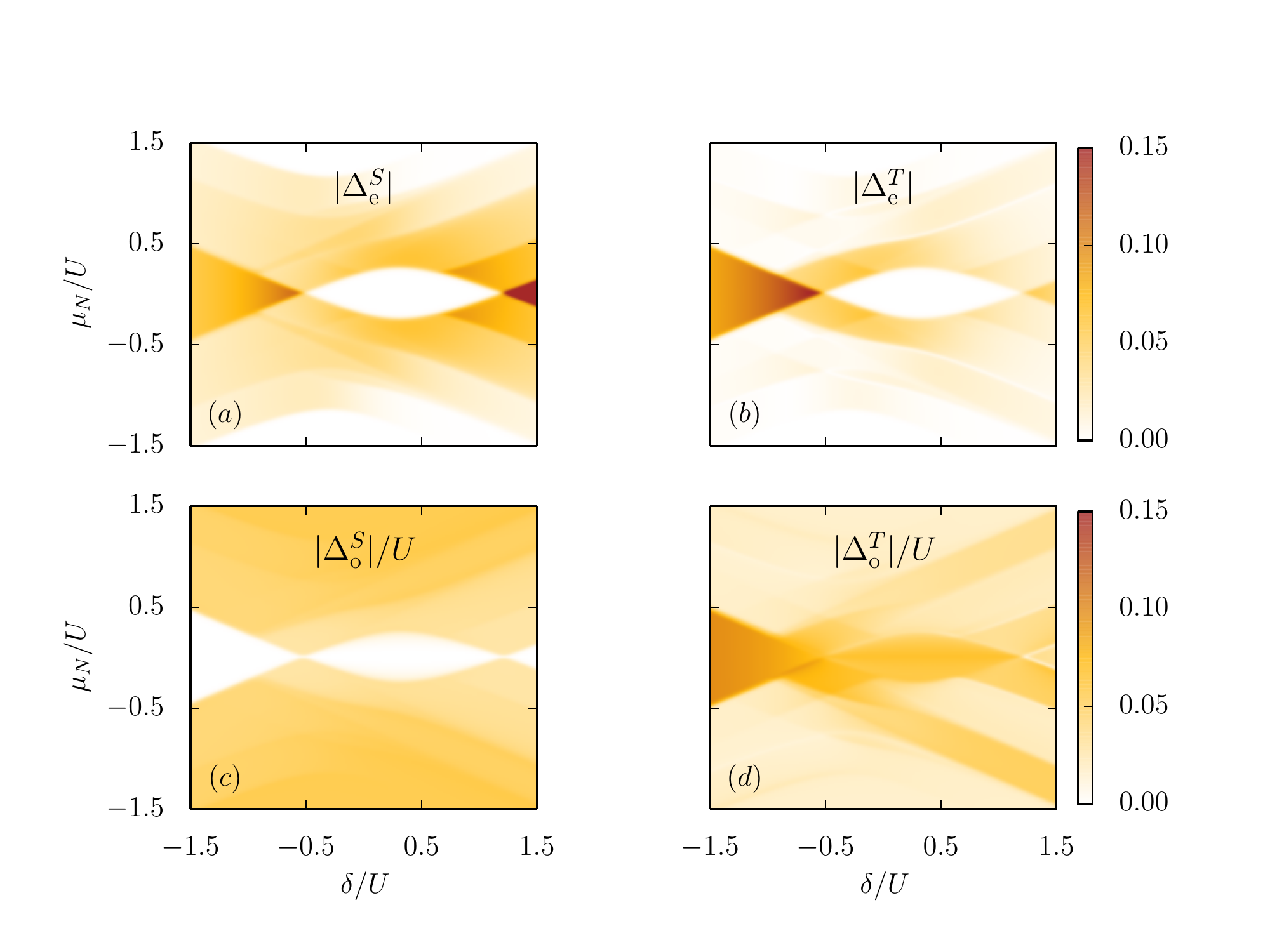}
\caption{\label{fig:order_deleps0} Absolute values of the superconducting order parameters as a function of $\delta$ and $\mu_\text{N}$ for zero detuning $\Delta \epsilon$ between the quantum-dot levels. The parameters are the same as in Fig.~3(a).}
\end{figure}

In Fig.~\ref{fig:order_deleps0} (a)-(d), we show the order parameters for the same system as in Fig.~3(a), i.e., for $\epsilon_\text{L}=\epsilon_\text{R}$ (no detuning between dot levels). 
We find that the gate and voltage dependence of the four order parameters strongly differ from each other. 
In particular, there are regions in which some of them vanish while others remain finite.
This is, e.g., the case in the Coulomb-blockade for small $|\delta|$ and $|\mu_\text{N}|$, where only the odd-frequency triplet order parameter is finite while all others vanish.
Similarly, for large bias voltage $|\mu_\text{N}|$, both the odd-frequencey singlet and the odd-frequency triplet order parameters survive while the even-frequency counterparts are suppressed.
With the help of Eqs.~(5a) and (5b), we can conclude that unconventional superconductivity is, in this case, generated by the terms involving a finite spin and a left-right asymmetry of the occupations in the DQD.
For large $|\delta|$ and small $|\mu_\text{N}|$, on the other hand, the odd-frequency singlet order parameter vanishes since in equilibrium the left-right symmetry is restored.

\hfill

\pagebreak

\subsection{B. Finite detuning between dot levels}

\begin{figure}[ht!]
	\includegraphics[width=0.98\columnwidth]{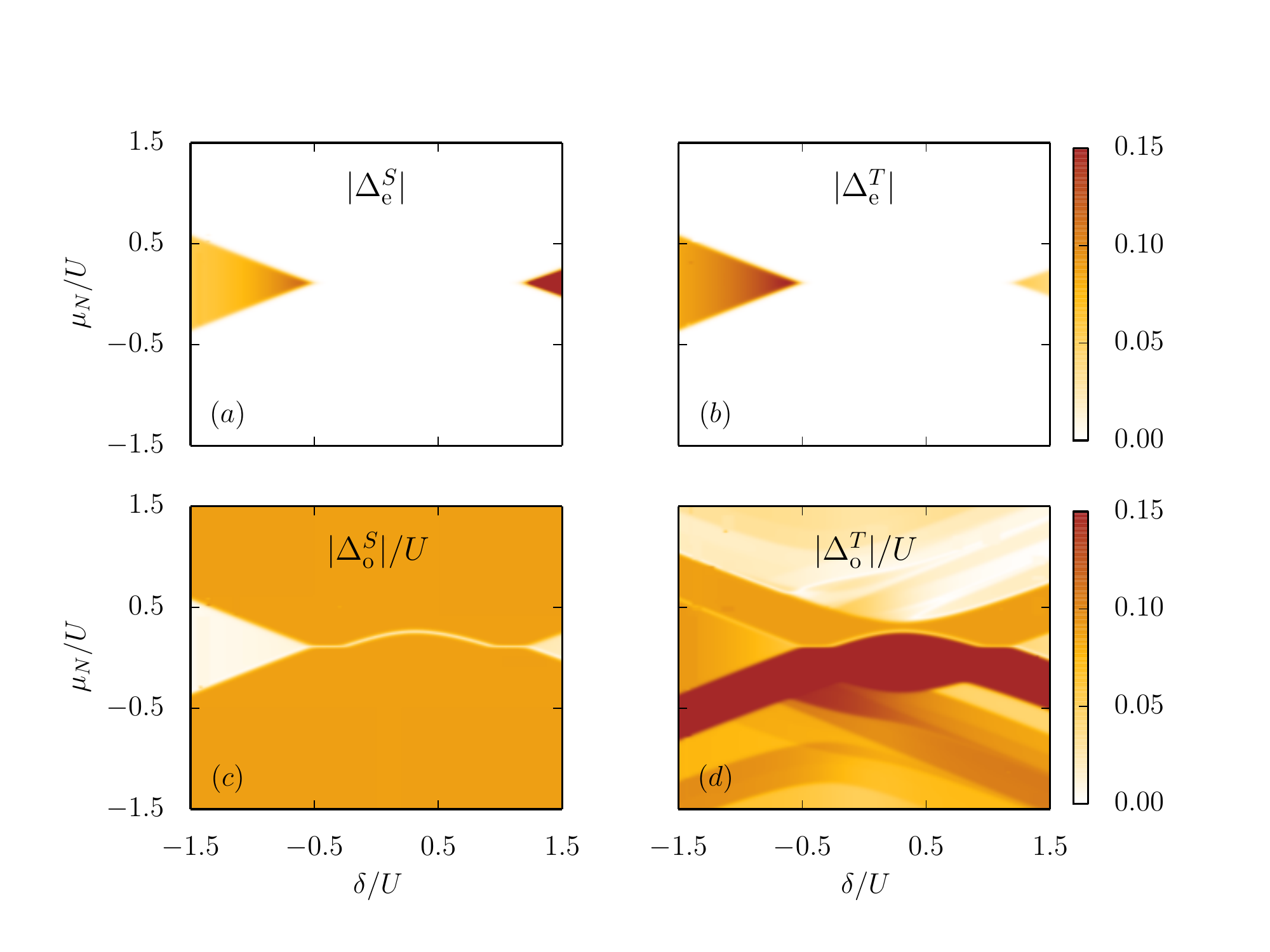}
\caption{\label{fig:order_deleps02} Same as Fig.~\ref{fig:order_deleps0} but for finite detuning $\Delta \epsilon=0.2U$ between the quantum-dot levels.}
\end{figure}

A finite detuning $\Delta \epsilon=\epsilon_\text{L}-\epsilon_\text{R}$ which is smaller than the interdot tunneling amplitude $t$ does not change the results qualitatively.
The situation becomes different for $\Delta\epsilon \gg t$.
In this case, an antisymmetrically applied bias voltage tends to favor a singly occupied state, which is incompatible with even-frequency order parameters, as clearly displayed in Fig.~\ref{fig:order_deleps02}.
(For a symmetrically applied bias voltage, as considered in Ref.~[42], the relative magnitude of $\Delta \epsilon$ and $t$ is not important.)
Odd-frequency order parameters, on the other hand, can still be finite due to the terms in the second line of Eqs.~(5a) and (5b).
The suppression of the even-frequency order parameters is lifted by increasing the interdot tunneling such that $t\gtrsim \Delta \epsilon$.
(We remark that breaking the left-right symmetry by $|{\bf B}_\text{L}| \neq |{\bf B}_\text{R}|$ while keeping $\epsilon_\text{L}=\epsilon_\text{R}$ leads to a similar suppression of even-frequency pairing. This was the motivation to choose a small angle $\gamma$ in Fig.~\ref{fig:Andreev_delB}(a) and (b).
For Fig.~\ref{fig:Andreev_delB}(c) and (d), i.e., symmetrically applied bias voltages this suppression is not an issue.)

\begin{thebibliography}{apsrev4-1}
\bibitem{footnote}
For all results, we use the full Hamiltonian without making use of the approximations leading to Eqs.~(\ref{i})-(\ref{iii}).
\end{thebibliography}
\end{document}